\documentclass[journal]{IEEEtran}
\IEEEoverridecommandlockouts
\ifCLASSINFOpdf

\else

\fi
\usepackage{amsmath}
\usepackage{graphicx}
\usepackage{diagbox}
\usepackage{caption}
\usepackage{subcaption}
\usepackage{epstopdf}
\usepackage{cite}
\usepackage{amsfonts}
\usepackage{amssymb}
\usepackage{algorithmic,algorithm}
\usepackage{textcomp} 
\usepackage{xcolor}
\usepackage{dsfont}

\begin{document}
 
\title{Analyzing {U}plink {G}rant-free {S}parse {C}ode {M}ultiple {A}ccess {S}ystem in {M}assive {I}o{T} {N}etworks}

\author{ Ke Lai, Jing Lei, Yansha Deng \IEEEmembership{Member,~IEEE}, Lei Wen, Gaojie Chen \IEEEmembership{Senior Member,~IEEE}
	\thanks{K. Lai,  J. Lei, L. Wen, are with Department of Communication Engineering, College of Electronic Science and Engineering, National University of Defence technology.  (e-mail: laike12@nudt.edu.cn; leijing@nudt.edu.cn; newton1108@126.com).}
	\thanks{Y. Deng,  is with Department of Engineering, King’s College London, London WC2R 2LS, U.K. (email:yansha.deng@kcl.ac.uk) }
	\thanks{G. Chen is with the Department of Engineering, University of Leicester,
		Leicester LE1 7RH, U.K. (e-mail: Gaojie.Chen@leicester.ac.uk). } }
\maketitle  

\begin{abstract}
Grant-free sparse code multiple access (GF-SCMA) is considered to be a promising multiple access candidate for future wireless networks. In this paper, we focus on characterizing the performance of uplink GF-SCMA schemes in a network with ubiquitous connections, such as the Internet of Things (IoT) networks. To provide a tractable approach to evaluate the performance of GF-SCMA, we first develop a theoretical model taking into account the property of multi-user detection (MUD) in the SCMA system. We then analyze the error rate performance of GF-SCMA in the case of codebook collision to investigate the reliability of GF-SCMA when reusing codebook in massive IoT networks. 
For performance evaluation, accurate approximations for both success probability and average symbol error probability (ASEP) are derived. To elaborate further, we utilize the analytical results to discuss the impact of codeword sparse degree in GF-SCMA.
After that, we conduct a comparative study between SCMA and its variant, dense code multiple access (DCMA), with GF transmission to offer insights into the effectiveness of these two schemes. This facilitates the GF-SCMA system design in practical implementation. Simulation results show that denser codebooks can help to support more UEs and increase the reliability of data transmission in a GF-SCMA network. Moreover, a higher success probability can be achieved by GF-SCMA with denser UE deployment at low detection thresholds since SCMA can achieve overloading gain.
\end{abstract}


\section{Introduction}

\IEEEPARstart{F}uture wireless communication systems are expected to achieve extended coverage, higher capacity, and massive connectivity with robust reliability to satisfy the growing requirements for information exchanges over mobile networks. The newly emerging application cases, such as the Internet of Things (IoT), corresponds to the situation that involves ubiquitous physical devices connected to a communication network through a wired or wireless channel \cite{2017Latency}.
 In IoT, machine-to-machine (M2M) communication supports an enormous number of random access (RA) user equipments (UEs), which are only sporadically active with small data transmission \cite{2017Wide}. Therefore, it is a vital issue to meet the demands of massive access with low signaling overload and delay in the future wireless communication \cite{2020A}.  

To support the congested traffic, massive connectivity, and vastly diverse quality-of-services (QoS) requirements in the massive IoT networks, a set of enabling technologies are introduced, in which non-orthogonal multiple access (NOMA) has received extensive research attention \cite{2018A}. NOMA is an emerging paradigm for future wireless network as it can achieve overloading gain. In contrast to orthogonal multiple access (OMA), NOMA serves multiple UEs with the same physical resource blocks (RBs), and thus allowing overloaded multiuser communications and enabling higher spectral efficiency. Based on the unique signatures that can be identified by the receiver, NOMA can be simply categorized into power-domain NOMA (PD-NOMA) and code-domain NOMA (CD-NOMA) \cite{2018A,2019Energy,2020Design,2019Buffer}. 
PD-NOMA advocates the superposition of two or more UEs who are assigned with different power levels over
the identical time-frequency resources \cite{2013Non}, while CD-NOMA employs carefully designed channel codes, interleavers, scrambling and spreading sequences to generate user-specific codewords \cite{2019Grant}.
The major focus of this work is on SCMA, which is a spreading based CD-NOMA scheme with relatively short spreading sequences. 

To reduce the costs, and yields a much longer battery life for the devices in massive IoT networks, it is necessary to effectively manage M2M communications with low signaling overhead and access delay. In the current long-term evolution (LTE) system, the legacy grant-based RA procedure is designed to provide reliable access to a small number of UEs, where the entity has to first go through a contention-based process over the physical random access channel (PRACH) to get aligned with the BSs. Although several modifications and improvements have been proposed \cite{2017Random,2018Optimization}, the type of LTE-based RA still faces a scarcity of the wireless resource. Thereby, grant-free (GF) transmission is proposed as a compelling alternative. In GF transmission, no further requests between BS and UEs should be made. Accordingly, an arrive-and-go transmission manner without extra signaling overhead is able to be realized.
As reported in \cite{2020A}, GF transmission with OMA causes severe collisions because of limited available resources, whereas NOMA can mitigate collisions as it serves different UEs over the same RBs with pre-determined user-specific signature patterns. Therefore, the combination of NOMA and GF transmission, which is denoted as GF-NOMA, is considered to be a promising technique to support uplink transmission in massive IoT networks.

Numerous researches have been conducted on CD-NOMA based networks and GF-CD-NOMA.  
An analytical expression of outage performance and optimal power allocation was derived in \cite{Chen2018Outage}. In addition,  closed-form of uplink SCMA error rate performance and sum rate  with randomly deployed users were presented in \cite{2018Performance} and \cite{Yang2017On}, respectively. As reported in \cite{Han2017A}, a resource scheduling scheme based on feedback for uplink GF-SCMA is illustrated to reduce the packet loss rate. An LTE-based SCMA RA, namely, SCMA applied RA (SARA) was proposed in \cite{2018SARA} to reduce the collision probability. 
In addition, the improvement of reliability in GF-CD-NOMA also receives extensive attention. A graph model for joint channel estimation and data detection was proposed for uplink GF-SCMA receiver, which can obtain significant performance gain compared to the disjoint detector \cite{2019Wei}. As reported in \cite{2017PoC}, a prototype of concept (PoC) of GF-SCMA system is realized. In \cite{2019Deep}, a deep learning aided GF-CD-NOMA scheme to meet the demands of high reliable and low latency in tactile IoT networks was proposed. In \cite{liu2021A}, a deep reinforcement learning framework for CD-NOMA was proposed to serve more UEs under strict latency constraint. 

Stochastic geometry has been regarded as a powerful tool to model and analyze mutual interference between transceivers in the wireless networks. Recently, to characterize the performance of multi-cell NOMA, stochastic geometry was applied as an analytical approach to account for inter-cell interferences in a practical network scenario. 
To evaluate the impact of the number of RBs, the success probability and area spectral efficiency (ASE) of SCMA is analyzed in \cite{2018Liu}, where an interference-limited transmission is considered for simplicity. In addition, only the users reusing the same codebook was identified as interference in that work. Different from \cite{2018Liu}, the interfering UEs were assumed to be allocated with the same pilots and codebooks in \cite{2019Evangelista}. The comparison between SCMA and OFDMA in device-to-device (D2D) communication was studied in \cite{Liu2017Modeling}. Note that, similar to \cite{2018Liu}, UEs with the same codebook were considered to be interferes in \cite{Liu2017Modeling}. The work in \cite{Abbas2019A} provided an analytical framework for single-cell CD-NOMA, where the authors mainly focused on the collisions caused by pilot sequences in GF transmission.  Furthermore, a novel analytic model for downlink NOMA in a single-cell is studied in \cite{Yue2018A}. In that work, PD-NOMA and CD-NOMA are analyzed in a unified framework. 

Based on the existing works, we can conclude that there are several aspects of limitation in the previous stochastic geometry based SCMA analysis: 1) 
Firstly, it should be noted that power control was of vital importance in uplink transmission, however, no explicit power control scheme has been considered in the aforementioned SCMA analytical works; 2) The multi-cell performance of SCMA analyzed in \cite{2018Liu,2019Evangelista,Liu2017Modeling} are mainly based on the signal-to-interference ratio (SIR), and thus the impact of noise is neglected; 3) Moreover, previous work on SCMA network mainly limited to model inter-cell interference as the reuse of codebook by different UEs, whereas the intra-cell interference is omitted and the impact of codebook collision is exaggerated. This is because SCMA can work under codebook collision, and the performance loss is affordable for the real system\cite{Lu2015Prototype}. Importantly, the property of MUD in SCMA, which is termed as message passing algorithm (MPA), is not considered in the existing works, i.e., the maximum number of UE that can be decoded by MPA is determined by the number of RB and codebook. 
 
In this paper, we figure out all above limitations and shed light on analyzing the success probability and error rate performance of uplink GF-SCMA with the truncated channel inversion power control \cite{2014ElSawy}. Note that, we take the property of MUD and SCMA codeword structure into account. To the best of our knowledge, such an analytical model of GF-SCMA has been rarely studied in the existing literature. 
The main contributions of this paper can be summarized as follows:

\begin{itemize}
	\item We analyze the success probability and ASE of the GF-SCMA system in a massive IoT scenario accounting for the property of MPA and truncated full channel inversion power control. 
\end{itemize}

\begin{itemize}
	\item Taking SCMA codebook design into account, the imapct of sparse degree of SCMA codeword and the number of codebooks are carefully analyzed based on the analytical model and simulation results. As a result, a comparative study between sparse and dense codebooks is presented.
\end{itemize}

\begin{itemize}
	\item The error rate performance results under codebook collision is also explored in this work, and an exact expression of average pairwise error probability (APEP) of GF-SCMA averaging with respect to transmit signals and UE deployment is derived.
\end{itemize}

\begin{itemize}
	\item We also conduct asymptotic analysis on the success probability and ASEP of GF-SCMA. By doing so, several meaningful results and insights can be obtained, which provides valuable guidance for the design of GF-SCMA system.  
	
\end{itemize}

The remainder of this paper is organized as follows. In Section II, the system model of GF-SCMA is briefly described. The success probability of uplink GF-SCMA in the network is discussed in Section III. In Section IV, the error rate performance of GF-SCMA under codebook collision is analyzed. The simulation results and discussions are presented in Section V. Finally, we conclude this paper in Section VI.

\section{System Model}\label{section.II}
In this section, we briefly introduce the main features of SCMA system, then the aspects of considered system model is presented in detail.
\subsection{Network Model and power control scheme}
In this paper, we consider a single-tier cellular network, where the BSs and the UEs are spatially distributed in $\mathbb{R}^2$ following two independent Homogeneous Poisson Point Processes (HPPPs) $\Phi_B$ and $\Phi_{UE}$ with intensities $\lambda_b$ and $\lambda_{ue}$, respectively. By Slivnyak's theorem \cite{2012Martin}, we condition on
having a $\text{BS}_o$ located at the origin, and studying the performance of the typical cell. Assuming that the distance between a UE and its serving BS is $R$ in the typical cell. Since $\Phi_B$ and $\Phi_{UE}$ are HPPPs, the distance between a UE and its serving BS follows a Rayleigh distribution with probability density function $f_R(x) = 2\pi\lambda x e^{-\pi\lambda x^2}$, where $x\geq 0$. 

In the uplink transmission, we consider a truncated channel inversion power control scheme. Conditioning on UE being served by $\text{BS}_o$, its transmission power is set as:
\begin{displaymath}
\mathcal{P}_{\text{UE}} = \left\{ \begin{array}{ll}
\rho R^{\eta\varepsilon} & \textrm{if $\rho R^{\eta\varepsilon} \leq \rho_m$}\\
\mathcal{P}_{bound} & \textrm{otherwise}
\end{array} \right.,\eqno(1)  \label{power_control}
\end{displaymath}
where $\rho$ represents the transmit power before power control, $\eta > 2$ is the path-loss exponent, $\varepsilon$ is the fractional power control factor, and $\rho_m$ is the maximum transmission power of a UE. When $\mathcal{P}_{bound} = 0$, and $\rho_m \neq \infty$, a fractional power control scheme with truncated transmission power can be obtained. In this case, the average signal power keep to be the same at the receiver. As such, if a UE moves closer to the desired BS, the power consumption of it should decrease to meet the requirement of a fixed average received power. Hence, lower cost and longer battery life are required in massive IoT networks. More importantly, such power control scheme can help to eliminate the near-far effect, and strike a good balance between the increasing interference and power consumption. 
In this work, for mathematical tractability, we assume that full path-loss inversion based power control is utilized, this indicates that $\varepsilon = 1$.
Besides, a fading channel that the channel response coefficient varies over RBs is considered. Thus, the channel response coefficient vector between a UE and the typical BS can be written as $\mathbf{h} = \left[h^{(1)}, \cdots, h^{(K)}\right]$, where all entries of $\mathbf{h}$ are independent and identically distributed complex Gaussian distribution $\mathcal{CN}(0,1)$. Without loss of generality, we assume that all BSs share the same $\eta$ and $\rho$.

\subsection{SCMA Transmission}
In this work, it is assumed that each UE transmits data to their served BS in  GF-SCMA manner. Moreover, we consider the uplink transmissions where the UEs are equipped with a single antenna. According to \cite{Hosein2013SCMA}, the SCMA codeword can be unified represented as a $K$-dimensional complex vector $\mathbf{c} = [c_1, \cdots, c_K]$, where $c_k \in \mathbb{C}$, $k \in \left\{1, \cdots, K\right\}$, and $\mathbf{c}$ is selected from a pre-defined codebook $\mathcal{C}$ with cardinality $M$, where $\mathcal{C} \subset \mathbb{C}^{K}$. As stated in \cite{Vameghestahbanati2019Multidimensional}, most of the well-performed SCMA codebooks with $M=4, 8$ and $16$ are designed according to the unimodular rule.
Accordingly, it is reasonable to assume that the considered codebook is of unimodular in this work.
Note that the codeword of conventional SCMA has a sparse structure, where the non-zero elements in a codeword are determined by the indicator matrix. Taking an SCMA system with $K=4$ physical resources and $L=6$ layers as an example, the indicator matrix can be represented as:

\setcounter{equation}{1} 
\begin{equation}
\begin{array}{l}
\boldsymbol{F}_{4\times 6 }=
\left[
\begin{matrix}
0&1&1&0&1&0\\
1&0&1&0&0&1\\
0&1&0&1&0&1\\
1&0&0&1&1&0 \label{indicator_matrix_1}
\end{matrix}
\right],
\end{array}
\end{equation}
where each row is associated to a specific RB and each column corresponds to the layer allocates to each UE. For brevity, we define sparse degree $d_s$, where $d_s \in \mathbb{N}^{+}$ and $(1 <d_s \leq K)$, as the number of non-zero elements in each column. Thereby, the design of SCMA can be interpreted as selecting $d_s$ positions out of $K$ to place $d_s$-dimensional constellation point. Then, the SCMA encoder can be defined as a mapping, $f: \mathbb{B}^{\log_2(M)} \to \mathcal{C}$, where $\mathbb{B}$ is the set of binary numbers. In a word, the encoder of SCMA makes a binary vector with $\log_2(M)$ bits map to a $K$-dimensional codeword selected from $\mathcal{C}$.
Note that,  the sparsity of the SCMA codewords is able to limit the number of users colliding over the same resource that in turn reduces the complexity of user detection. Thus, MPA can be implemented to the detection of a SCMA system with low complexity and high reliability. For MPA, it is noteworthy that the data of UEs is recovered in a parallel manner. As $L = \frac{K!}{d_s!\left(K-d_s\right)!}$ layers are multiplexed over $K$ physical resource elements or equivalently orthogonal frequency-division multiple access (OFDMA) tones. Then, if $T$ OFDMA tones are available for transmission,  $J =LT/K$ distinguished codebooks can be utilized by UEs in a cell.

  \setcounter{equation}{2}  
\begin{figure*}[t!]
	\begin{equation}
	\begin{aligned}	 
	\mathbf y_o &= \sum\limits_{j\in\mathcal{U}}\sqrt{\rho} \mathbf{c}_j\text{diag}(\mathbf{h}_j)\mathds{1}\left\{R_j\leq\left(\dfrac{\rho_m}{\rho}\right)^{1/\eta}\right\}  +  \underbrace{\sum\limits_{u\in\mathcal{U}_{in}\backslash \mathcal{U}}\sqrt{\rho }\mathbf{c}_u\text{diag}(\mathbf{h}_u)\mathds{1}\left\{R_u\leq\left(\dfrac{\rho_m}{\rho}\right)^{1/\eta}\right\}\mathds{1}\left\{\left\vert\mathcal{U}\right\vert > J\right\}}_{\mathcal{I}_{intra}} \\
	& \quad\quad\quad\quad\quad\quad\quad + \underbrace{\sum\limits_{\text{UE}_i \in \Phi_U\backslash \mathcal{U}_{in} }\sqrt{\rho R_i^{\eta}} D_i^{-b}\mathbf{c}_i\text{diag}(\mathbf{h}_i)\mathds{1}\left\{R_i\leq\left(\dfrac{\rho_m}{\rho}\right)^{1/\eta}\right\}\mathds{1}\left\{D_i > R_i\right\}}_{\mathcal{I}_{inter}} + \mathbf{z}_o. 	
	\label{superposed_sig}
	\end{aligned}
	\end{equation}
	\hrulefill
\end{figure*}

\subsection{Contention transmission unit (CTU) for GF-SCMA}

\begin{figure}[t!]
	\centering	
	\renewcommand{\captionfont}{\small } \renewcommand{\captionlabelfont}{\small} \centering \includegraphics[height=2.0in, width=2.8in]{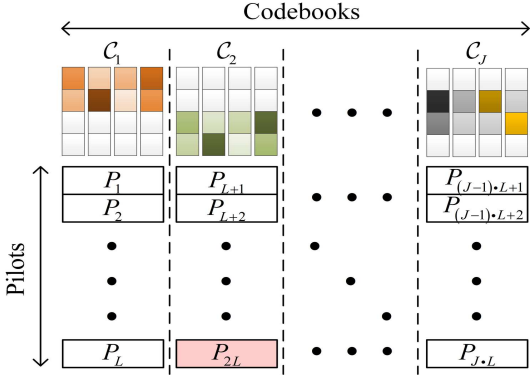}\caption{Illustration of CTUs.}	
	\label{fig_CTU}
	
\end{figure}

For GF transmission, the basic resource for GF-SCMA system is named as CTU. It is defined as a combination of time, frequency, SCMA codebook and pilot \cite{2014Uplink}. As shown in Fig. \ref{fig_CTU}, over a time-frequency resource, there are $J$ groups, each group is assigned to a codebook $\mathcal{C}_j$ ($0\leq j \leq J$), and pilot $P_{\left(j-1\right)\cdot L + l}$ ($0\leq l \leq L$), where $L$ pilots in each group generates from the same Zadoff-Chu sequence with different cyclic-shifts. By doing so, a pilot pool with $JL$ unique pilots can be defined.
In a time slot, each RA UE randomly selects a pilot from the pilot pool and the corresponding codebook from the CTU. For instance, as can be seen from Fig. \ref{fig_CTU}, if the pilot that is indexed by $2L$ is selected, then codebook $\mathcal{C}_2$ is utilized to transmit data. Moreover, it is also noteworthy that one codebook can map to $L$ pilots, which indicates that even though the codebook is reused by many CTUs in a GF-SCMA network, they can be identified as long as CTUs are of different pilots. Considering the sporadic communication in massive IoT networks, where the active UEs only constitute a small subset of all potential IoT UEs. In this paper, UEs that transmit data are named as active UEs, while the other UEs are deemed to be inactive. Note that, the active probability is defined as $p_a\in[0,1]$, which means that only a fraction $p_a$ of UEs is assumed to be activated and transmitting data at the considered time slot.

Therefore, in a transmission period of GF-SCMA, BSs first perform uplink radio resource control (RRC) to make sure that each active UE is allocated with channel resources to enable GF transmission. After that, 
a UE, which has data to transmit, conveys their data in a GF-SCMA manner by using a randomly selected CTU. If different pilots are utilized by active UEs, then it is assumed that the RA UE could be identified successfully. By contrast, once the same pilot is selected by different UEs, a pilot collision occurs and the data transmission is deemed to be failed. 



\subsection{Signal to Noise plus Interference Ratio (SINR)}  \label{Section-II-D}
\begin{figure}[t!]
	\centering	
	\renewcommand{\captionfont}{\small } \renewcommand{\captionlabelfont}{\small} \centering \includegraphics[height=2.5in, width=3.8in]{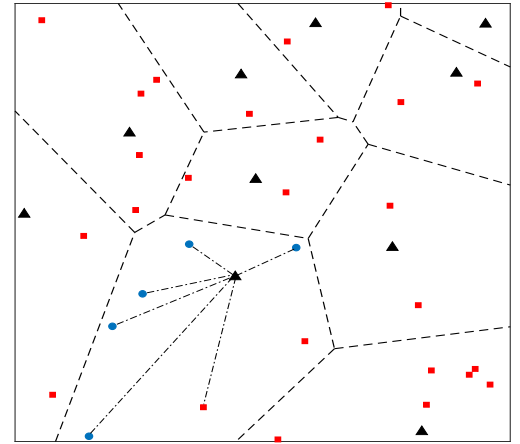}\caption{A realization of the GF-SCMA network, where $J = 6, K = 4$. The triangle $``\blacktriangle"$ denotes the BS, square $``\blacksquare"$ is the UEs with the same pilot, and circle $``\bullet"$ represents the UEs that have different CTUs.}
	\label{fig_network_plot}	
\end{figure}

	Different from the existing works in  \cite{2018Liu,2019Evangelista,Liu2017Modeling} that only considered imposing interference on one typical UE on the condition that codebook collision occurs, this paper considers the property of CTU and MPA. As a result, the SINR of GF-SCMA is modeled based on the following two facts: 
\begin{itemize}
	\item[1)] In contrast to the conventional OMA and PD-NOMA that recover data from multi-user in a serial manner; hence, the data is detected UE-by-UE. For GF-SCMA, $J$ UEs' data are decoded simultaneously by using MPA. As such, the success probability of GF-SCMA is not just determined by a typical UE but all the UEs that are involved in MPA. However, MPA cannot work under the condition that the practical overloading factor is higher than $L/K$ when the same number of RB is utilized. Thereby, at most $J$ UEs with different pilots can be decoded at the same time in a typical cell.
\end{itemize}
\begin{itemize}
	\item[2)] If UEs in the network select CTUs with different pilots for transmission, the UE can be identified and the data can be decoded even though codebook collision occurs \cite{Lu2015Prototype}. However, if CTUs with the same pilot intend to transmit data at the same time, not only the UE cannot be identified, but also the data can not be recovered or canceled since the channel condition is unknown. As such, they should be treated as interference.
	
\end{itemize}
For ease understanding, an example of GF-SCMA network considered in this work is presented in Fig. \ref{fig_network_plot}. As can be seen in Fig. \ref{fig_network_plot}, in a typical cell, $J = 6$ UEs with different CTUs intend to communicate with the BS. Assuming these $J$ UEs possess distinguished CTUs, and they can process MPA perfectly at the receiver. Accordingly, the interference comes from the UEs that using the same pilot as one of the $J$ UEs, which is denoted by ``square" in the figure.

From above discussions and accounting for the power control scheme in (1), the superimposed signal received at the typical BS can be generally expressed as \eqref{superposed_sig}, 
where $\mathcal{U}_{in}$ is the set of active UEs with pilot collisions in a specific Vironoi cell, $\mathcal{U}$ is the set of UEs that can be decoded by using MPA, and they are identified by the subscript ``\emph{j}''\footnote{In general, we have $\vert \mathcal{U}\vert \leq J$, and if $\vert \mathcal{U}\vert<J$, $J - \vert \mathcal{U}\vert$ all-zero codewords should be involved in MPA as reported in \cite{2014Uplink}}. Note that the cardinality of $\mathcal{U}$ can be formulated as
\begin{displaymath}
\vert\mathcal{U}\vert = \left\{\begin{array}{ll}
\vert\mathcal{U}_{in}\vert & \textrm{if $\vert\mathcal{U}_{in}\vert < J$}\\
J & \textrm{if $\vert\mathcal{U}_{in}\vert \geq J$}
\end{array}. \right. \eqno(4)\label{Num_UE}
\end{displaymath}
In addition, $\mathbf{z}_o$ is the noise vector that consists of complex Gaussian random variables with distribution $\mathcal{CN}(0,\sigma^2)$. In addition, $D_i$ is the distance between $\text{BS}_o \in\Phi_B$ and $\text{UE}_i \in\Phi_{UE}$, and $\mathds{1}\left\{\cdot\right\}$ denotes an indicator function takes value 1 if the condition $\mathds{1}\left\{\cdot\right\}$ is satisfied.
According to \eqref{superposed_sig}, a general expression for the SINR of $\text{BS}_o$ can be formulated as
\setcounter{equation}{4}    
\begin{equation}
\text{SINR} = \dfrac{\sum_{j\in\mathcal{U}} P_j \mathcal{G}_j}{\mathcal{I}_{inter} + \mathcal{I}_{intra} + \sigma^2}, \label{SINR}
\end{equation}
where
$\mathcal{G}_j=\sum_{s=1}^{d_s} \vert c_{j,s} \vert^2 \vert h_j^{(s)}\vert^2 = \sum_{s=1}^{d_s}\vert g_j^{(s)}\vert^2$, in which $g_j^{(s)}$ is the $s$th entry of $\mathbf{c}_j\text{diag}(\mathbf{h}_j) \in \mathbb{C}^{1\times K}$. $\mathcal{I}_{intra}$ and $\mathcal{I}_{inter}$ denote the intra- and inter-cell interference, respectively. Furthermore, it should be noted that $\mathcal{G}_u= \sum_{s=1}^{d_s}\vert g_u^{(s)}\vert^2$ and $\mathcal{G}_i= \sum_{s=1}^{d_s}\vert g_i^{(s)}\vert^2$ have the same formulation as $\mathcal{G}_j$. As the channel response coefficient follows complex Gaussian distribution, and thus $\mathcal{G}_{\Xi} \sim \text{Gamma}(d_s,1)$, in which $\Xi \in \{j,u,i\}$. For simplicity, we also define $\mathbf{H}_{\Xi} = \text{diag}(\mathbf{h}_{\Xi}) $,
$b = \frac{\eta}{2}$, and $P_i = \rho R_i^{\eta}$ in the following of this paper.

Moreover, it can be also observed from \eqref{superposed_sig} that useful signals are superimposed by at most $J$ RA UEs in a typical cell.  As a result, the intra-cell interference exists only when the number of active UEs is larger than that of supported UEs for a GF-SCMA system, and thus the number of intra-cell interferes equals to $\left\vert\mathcal{U}_{in}\right\vert - J$. Remind that this is because MPA is unable to decode SCMA codewords once reaching the maximum number of supported UEs. As such, an important assumption should be made in this work, which is demonstrated as follows.
As depicted in Fig. \ref{fig_network_plot}, taking account for a network where the intensity of UE is $\lambda_u^{'} = p_a\lambda_{ue}/JL$. We consider a typical cell in which randomly selected $\vert\mathcal{U}\vert$ UEs are assumed to have distinguished CTUs. In this case, we assume that the remaining UEs have the same pilot as one of the $\vert\mathcal{U}\vert$ UEs, such that the realistic intensity of interferes $\lambda_u$ is approximated as the intensity of UE that has the same pilot, i.e., $\lambda_u \approx \lambda_u^{'}$. The feasibility of such an approximation is verified in Section. \ref{Simulation}.

\section{SINR analysis} 
To evaluate the success probability performance of the GF RA with SCMA, in this section, we first derive the success probability of GF-SCMA, then some important properties and interesting insights are presented. 

\subsection{Success Probability Analysis} \label{SP_ana}
From the previous discussion, it is important to note that the key factors influencing the success probability can be outlined as follows:  1) Only RA UEs without experiencing pilot collisions have the opportunity for BS to recover the transmitted data; hence, the interference come from UEs with the same pilot;
2) The signal of RA UEs transmitting with GF-SCMA cannot be recognized by its serving BS, since its lower received SINR than the SINR threshold $\gamma_{\text{th}}$, such that outage occurs; 3) As MPA is utilized to recover the data of each UE in GF-SCMA, the interference comes from intra- and inter-cell superimpose on the signals of at most $J$ UEs that are multiplexed by $T$ RBs.   


  
In the sequel, the mathematical expression of success probability is derived by taking the above influencing factors and the approximation in Sec. \ref{Section-II-D} into account. 
To apply Gil-Palaez inversion theorem to calculate success probability\cite{1951gil,Deng2019SINR}, we first consider the characteristic function (CF) of the interference according to \eqref{superposed_sig}, and the definition of SINR in \eqref{SINR}. 
Here, we assume that the data of UEs belong to $\mathcal{U}$ can be perfectly decoded; hence, we have following lemma. 

\noindent \textbf{Lemma 1.} \emph{The CFs of inter-cell and intra-cell  interference for SCMA transmission are given as
}
\newcounter{TempEqCod}                         
\setcounter{TempEqCod}{\value{equation}} 
\setcounter{equation}{5} 
\begin{equation}
\Phi_{\mathcal{I}_{inter}} =\exp\left\{\beta c\left[1-\left(1+\frac{\alpha}{\pi\lambda_b}\right)\exp\left(-\alpha\right)\right]\mathcal{M}_{\omega}
\right\} \label{CF_inter}
\end{equation}
\setcounter{equation}{\value{TempEqCod}}
\emph{and}
\setcounter{equation}{6}    

\begin{equation}
\begin{aligned}
\Phi_{\mathcal{I}_{intra}} = 
\mathcal{K}_{\omega}^{-\left(J-1\right)} \Bigg\{\mathbb{P}&\left\{\mathcal{U}_{in} = u\right\} \left[\mathcal{K}_{\omega}^{\left(J-1\right)} - \mathcal{K}_{\omega}\right] \\
 &\quad+  \left(1+\beta\left(1-\mathcal{K}_{\omega}\right)^{-c-1}\right)\Bigg\},  \label{CF_intra}
 \end{aligned}
\end{equation}
\emph{respectively, where $\mathcal{M}_{\omega} = 1-{}_{2}F_{1}(-\frac{1}{b},d_s;1-\frac{1}{b};j\omega \rho)$\footnote{Unless otherwise stated, ${}_{p}F_{q}$, where $p$ and $q$ are two nonnegative integers, denotes the generalized generalized hypergeometric functions in this paper.}, $\alpha = \pi\lambda_b\left(\rho_m/\rho\right)^{1/b}$, $\beta = (\mathcal{O}_p\lambda_u/c\lambda_b)$, and $c = 3.575$ is a constant related to the approximate probability mass function of the HPPP Voronoi cell. $\mathcal{K}_{\omega} = \mathbb{E}\left[\exp\left(j\omega\mathcal{G}_{\Xi}\right)\right]=(1-j\omega \rho)^{-d_s}$, and $\mathcal{O}_p = exp\left(-\alpha\right)$ is the truncation outage probability}.

\emph{Proof}. See Appendix. \ref{Lemma 1}.

Note that, to provide a general analytical framework of GF-SCMA, the inter- and intra-cell interference are both considered. Remind that the interference comes from the reuse of pilot in CTUs.
According to the discussion in Sec. \ref{section.II}, we define a success transmission as the nonexistence of pilot collision and $\text{SINR}$ at the receiver satisfies  $\text{SINR}\geq \gamma_{\text{th}}$, and thus the following theorem can be given:

\noindent \textbf{Theorem 1.} \emph{The success probability of UEs belong to $\mathcal{U}$ in the uplink GF-SCMA system is derived as}
\begin{equation}
P_{\text{suc}} = \dfrac{1}{2} - \dfrac{1}{\pi}\int_{0}^{\infty}\Im\left\{\mathcal{S}_{\omega}\Phi_{\mathcal{I}_{intra}}\Phi_{\mathcal{I}_{inter}}\exp\left(j\omega\sigma^2\right) \right\}\dfrac{\text{d}\omega}{\omega},  \label{Theo_1}
\end{equation}
\emph{where $\Im\left\{\cdot\right\}$ denotes the imaginary part of a function, and $\mathcal{S}_{\omega}$ is a signal related function that is given as}
\begin{equation}
	\begin{aligned}
\mathcal{S}_{\omega} = &\exp\left(-\alpha\right) + \left[1-\exp\left(-\alpha\right)\right] \Bigg\{1 - \sum_{u=0}^{J-1}\mathbb{P}\left\{\mathcal{U} = u\right\} \times\\
&\left[\left(1+j\omega \rho\frac{1}{\gamma_{\text{th}}}\right)^{-d_su}-\left(1+j\omega \rho\frac{1}{\gamma_{\text{th}}}\right)^{-Jd_s}\right]\Bigg\} ,
\end{aligned}
\end{equation}

\emph{Proof}. See Appendix. \ref{Theorem 1}.

In Theorem 1, we can observe that when considering full channel inversion power control without truncation, $\Phi_{\mathcal{I}_{inter}}$, $\Phi_{\mathcal{I}_{intra}}$ and $\mathcal{S}_{\omega}$ can be further simplified by substituting $\mathcal{O}_p \to 1$ and $\exp\left(-\alpha\right) \to 0$. This follows from the fact that  $\rho_m \to \infty$, such that $\alpha \to \infty$.
 In the following subsection, we will provide some key properties of the success probability, which will be verified in Section. \ref{Section_V} via numerical results.

\subsection{Key properties of success probability} \label{Section.III-C}
In SCMA, the codebook design method is the main characteristic that distinguishes it from other multiple access schemes. In this regard, it is necessary to study the influence of $d_s$ on the success probability. To this end, we provide the following two corollaries.

\noindent \textbf{Corollary 1.} \emph{Considering a interference-limited GF-SCMA network with $\sigma = 0$ and fixed $\lambda_u$, $\lambda_b$. In addition, it is assumed that the number of codebook is large enough to support all the intra-cell UEs whatever $\lambda_u$ and $\lambda_b$ vary. Then, it can be proved that $P_{\text{suc}}$ asymptotically scales with $\sqrt{\xi}\exp\left(-(\lambda_u/\lambda_b)^2\xi\right)$, where $\xi = \frac{1}{1+1/d_s}$.}

\emph{Proof}: See Appendix \ref{Corollary 1}.

It can be seen that $\sqrt{\xi}$ and $\exp\left(-(\lambda_u/\lambda_b)^2\xi\right)$ have different monotonicity.
 Nevertheless, as is shown in \eqref{Success_appr}, when $\lambda_u/\lambda_b$ is relatively low, it can be inferred from Corollary 1 that codewords with relatively large $d_s$ is more probable to have superiority over the sparser ones, especially in the region of low and medium detection threshold. To elaborate further, the following conclusions are given based on Corollary 1.
 
\noindent \textbf{Corollary 2.} \emph{ Denote $P_\text{suc}\left(d_s\right)$ as the success probability of GF-SCMA with codebook sparsity degree $d_s$,  then we have following two formulas}
\begin{equation}
\lim\limits_{d_s\to \infty} \left[P_{\text{suc}}\left(d_s + d\right) - P_{\text{suc}}\left(d_s\right) \right]= 0, \label{d_s to infty}
\end{equation}
\emph{and}
\begin{equation}
P_{\text{suc}}\left(d_s + d\right)-P_{\text{suc}}\left(d_s \right) \propto \dfrac{1}{\gamma_{\text{th}}} \label{P_suc differential}
\end{equation}
\emph{where $d \in \mathbb{N}^{+}$.}

\emph{Proof}:  We first prove \eqref{d_s to infty}. Based on the proof of Corollary 1, when $d_s \to \infty$, $\mathcal{S}_{\omega}\to \left(1+\beta\right)^{-c-1}$, such that $P_{\text{suc}}$ scales with $\xi^{1/2}$, which is uncorrelated to other parameters. After that, when $d_s \to \infty$, we have $\xi^{1/2} \to 1$, and thus \eqref{d_s to infty} is satisfied. As for \eqref{P_suc differential}, it can be directly inspected by substituting $d$ and $d_s + d$ into \eqref{Success_appr}. This concludes the proof.

According to the results in these two corollaries, we can have the following implications: 1) Although the success probability of GF-SCMA benefits from the increase of sparsity degree $d_s$ at low and medium $\gamma_{\text{th}}$ region when a large number of UE is deployed, no performance gain can be enjoyed by making the codebook denser\footnote{In general, there are two approaches to increase $d_s$ in SCMA without changing $J$. One is to adjust $K$ and $T$ as the increase of $d_s$, the other is to adopt novel codebook design rules, such as density code multiple access (DCMA), which is recently proposed in \cite{liu2020sparse}, and the sparse degree $d_s = K$.}. This property is important for the design of codebook with large $J$ and $K$; 2) According to \eqref{P_suc differential}, the performance gap between the sparse and dense codebooks shrinks as the increase of detection threshold. This property means that the codebooks of GF-SCMA with larger $d_s$ are able to achieve higher success probability in the region of low and medium $\gamma_{\text{th}}$, which is useful in the design of practical system; 3) It can be seen from \eqref{Success_appr}, as the increase of $\lambda_u/\lambda_b$, the impact of $d_s$ on $P_{\text{suc}}$ is negligible, and thus the gap between $P_{\text{suc}}$ and $P_{\text{suc}}\left(d_s + d\right)$ becomes marginal. In a word, it can be inferred that a denser deployment of UEs reduces the influence of codebook design in the GF-SCMA system.
Besides, based on Theorem 1, another interesting property can be obtained, which is demonstrated as follows,

\noindent \textbf{Corollary 3.} \emph{Considering an interference-limited scenario in a GF-SCMA network with fixed $\lambda_b$, and denote $P_{\text{suc}}\left(\lambda_u\right)$ as the success probability with UE's intensity $\lambda_u$. Then, $P_{\text{suc}}\left(\lambda_u\right) - P_{\text{suc}}(\lambda_u^{'}) \geq 0$ can be satisfied in the condition of  $\lambda_u > \lambda_u^{'}$. Moreover, as $\lambda_u$ increases, lower SINR threshold $\gamma_{\text{th}}$ should be taken to make $P_{\text{suc}}\left(\lambda_u\right) = P_{\text{suc}}(\lambda_u^{'})$.}

\emph{Proof}: Similar to the proof in Corollary 1, according to the approximation in \eqref{Success_appr}, it can be inferred that  $P_{\text{suc}}\left(\lambda_u\right) - P_{\text{suc}}(\lambda_u^{'})$ mainly depends on the function $f(x) = \xi^{1/2}x - x/\gamma_{\text{th}}\exp(-\xi x^2)$, where $x$ is a variable proportional to $\lambda_u$ here. By studying the first-order derivative of $f(x)$, i.e., $f^{'}(x) = \xi^{1/2} - 1/\gamma_{\text{th}}\exp(-\xi x^2)(1-2\xi x^2)$, we can observe that the monotonicity of $f(x)$ changes at a stationary point $x>0$. On the other hand, by direct inspection of the equation $f^{'}(x) = 0$, the stationary point $x$ increases by decreasing $\gamma_{\text{th}}$. This completes the proof. 
 
This result actually is not quite straightforward because using a fixed number of BSs to support more UEs should reduce the success probability intuitively. An intuitive explanation of this property is that even though more UEs are deployed in the network when $\lambda_u$ is relatively high, in this case, $\vert\mathcal{U}\vert$ approximately equals to $J$ on average, and thus the aggregate interference of UEs belong to $\mathcal{U}$ is also high. For the case when $\lambda_u$ is low, although less interference is undergone, the aggregate power of superimposed signal in $\mathcal{U}$ is also low. Since $\vert \mathcal{U} \vert$ is more probable to be small with low $\lambda_u$, and thus leading to a fail transmission in the region of low $\gamma_{\text{th}}$. 

\subsection{ASE analysis}
Furthermore, after attaining success probability, the ASE of GF-SCMA network can be also derived by using 
\begin{equation}
\mathcal{E}_A= \lambda_u\bar{\mathcal{U}}P_{\text{suc}}\log_2\left(1 + \gamma_{\text{th}}\right), \label{ASE}
\end{equation}
where $\bar{\mathcal{U}}$ is the mean number of served UEs, which can be further written as:
\begin{equation}
\bar{\mathcal{U}} = 1 + \sum_{u = 1}^{J-1}\mathbb{P}\left\{\left\vert\mathcal{U}_{in}\right\vert = u\right\}u + \sum_{u = J}^{\infty}\mathbb{P}\left\{\left\vert\mathcal{U}_{in}\right\vert = u\right\}(J-1). 
\end{equation} 
It should be noted that the ASE and success probability depict two different aspects of a communication system. A higher success probability means that the UEs can transmit data more reliable so that they enjoy a better quality of experience, whereas ASE denotes the spatial reuse efficiency, and thus more UEs can be supported in a network with a higher ASE.

\section{Error Rate performance analysis} \label{error_assup}
In this section, we  investigate the error rate performance on the condition that the same codebook is utilized by the interferes who have suffered from pilot collision. 
\subsection{Uplink Interference Model for Error Performance analysis}
It is known that the codewords of SCMA are well designed according to the method presented in \cite{Taherzadeh2014SCMA}, and thus the UEs that use different codebooks can recover the data with high reliability. Even though the effect of reusing the same codebook for different UEs can be mitigated due to the application of advanced MUD (i.e. MPA) and the reciprocity of wireless channel in GF-SCMA. However, it also causes error rate degradation \cite{YangSCMA,Lu2015Prototype,2017PoC}. Thereby, the codebook collision is deemed to be the main influencing factor of the error rate performance of GF-SCMA, especially in a large-scale network. Motivated by such facts, we examine the error rate performance of GF-SCMA in the case of codebook collision. Although errors occur even though different codebooks are utilized in link-level simulations \cite{Hosein2013SCMA}, the error rate performance deteriorates significantly when the same codebooks are utilized. Therefore, we can conclude that the error rate performance of GF-SCMA is mainly determined by codebook collisions in the considering model. 

Similar to the analysis in Section. \ref{SP_ana}, it is assumed that MPA can perfectly cancel the interference when $\vert \mathcal{U}\vert \leq J$. Besides, we only consider the case that the same codebook is reused at most once for tractability. That is, we end up with a maximum of one collided UE per BS.
The rationale behind such an assumption follows from the fact that
each cell shares the same codeword pool in GF-SCMA, and user-specific codebooks are assigned to the active UEs in each cell. Furthermore, taking into account the sporadic communication in massive IoT networks, the probability of reusing the same codebook more than once by different UEs is very low. This can be explained by the fact that the number of codebook $J$ is sufficient to support all UEs in a cell in most cases. Therefore, each BS is assumed to serve at most one UE using the same codebook. In a word, the derived results in this section intends to reveal the error rate performance when CTU collision occurs in a GF-SCMA system, this can help to examine the robustness of GF-SCMA under overloading condition.

\subsection{Pairwise Error Probability (PEP) analysis}
It is acknowledged that various MPA based algorithms can be applied to the detection of SCMA. Therefore, to provide a unified analytical model for the error performance, an optimal receiver, i.e., maximum likelihood detector (MLD) is considered in this work. In addition, we intend to compute the average symbol error probability (ASEP) from APEP, which has been utilized to evaluate the SCMA error rate performance in the single-cell with a fixed number of randomly deployed users in \cite{2018Performance,Bao2017Joint,2020Codeword}. 

For GF-SCMA, the MLD with perfect channel state information (CSI) at the receiver's side can be formulated as
	\begin{equation}
	\hat{\mathbf{c}_o} = \arg\min\limits_{\mathbf{c}_o \in \mathcal{C}} \left\{\left\Vert\mathbf{y}_o - \sqrt{\rho}\mathbf{c}_o\mathbf{H}_o \right\Vert^2 \right\}, 
    \end{equation} 
 where $\Vert\cdot\Vert$ represents the Euclidean distance. According to the definition of PEP, the PEP of SCMA codeword in a GF-SCMA network can be written as
	\begin{equation}
	\begin{aligned}
	P(&\tilde{\mathbf{c}_o} \to \mathbf{c}_o)\\
	 &= \text{Pr}\left\{\left\Vert \mathbf{y}_o - \sqrt{\rho}\tilde{\mathbf{c}_o}\mathbf{H}_o \right\Vert^2<\left\Vert \mathbf{y}_o - \sqrt{\rho}\mathbf{c}_o\mathbf{H}_o \right\Vert^2 \right\}\\
	 &\stackrel{(a)}{=} \text{Pr}\Big\{ 2\sqrt{\rho} \Re\left\{ \left(\sqrt{\rho}\mathbf{c}_o\mathbf{H}_o + \mathcal{V}\right) \left(\mathbf{\Delta}_o\mathbf{H}_o\right )^{\dagger} \right\} + \\
	&\quad\quad\rho\left(\Vert \tilde{\mathbf{c}_o}\mathbf{H}_o\Vert^2 - \Vert \mathbf{c}_o\mathbf{H}_o\Vert^2\right) < 0 \Big\}\\
	& = \text{Pr}\left\{\rho\Vert \mathbf{\Delta}_o\mathbf{H}_o \Vert^2 + 2\sqrt{\rho}\Re \left\{\mathcal{V}\left(\mathbf{\Delta}_o\mathbf{H}_o\right)^{\dagger}\right\} < 0 \right\} \\
	& \stackrel{(b)}{=} \text{Pr}\left\{\mathcal{Z}\left(\mathbf{\Delta}_o\mathbf{H}_o\right)^{\dagger} <  -\dfrac{\sqrt{\rho}}{2}\Vert \mathbf{\Delta}_o\mathbf{H}_o \Vert^2 \right\} \\
	&=  F_{\bar{\mathcal{V}}}\left(-\dfrac{\sqrt{\rho}}{2}\Vert \mathbf{\Delta}_o\mathbf{H}_o \Vert^2\right)
	\label{PEP}
	\end{aligned}
	\end{equation} 
where ($\mathnormal{a}$) follows from defining $\mathcal{V} = \mathcal{I}_{inter} + \mathbf{z}_o$,  $\mathbf{\Delta}_o = \mathbf{c}_o - \tilde{\mathbf{c}_o}$, and ($\mathnormal{b}$) holds because $\mathcal{V}$ is a circularly symmetric random variable (RV), $\Re\left\{\cdot\right\}$ denotes the real part of a complex number, and $\dagger$ is the conjugate transpose of a complex vector. $F_{\bar{\mathcal{V}}}$ is the cumulative density function (CDF) of $\bar{\mathcal{V}} = \mathcal{V}\left(\mathbf{\Delta}_o\mathbf{H}_o\right)^{\dagger}$. For simplicity, we define $\mathcal{Z} = \mathbf{z}_o \left(\mathbf{\Delta}_o\mathbf{H}_o\right)^{\dagger}$, and $\mathcal{I} = \mathcal{I}_{inter} \left(\mathbf{\Delta}_o\mathbf{H}_o\right)^{\dagger}$, such that  $\Re\left\{\mathcal{Z}\right\}$ has the same distribution as $\mathcal{V}$.
By using Gil-Palaez inversion theorem, $F_{\bar{\mathcal{V}}}$ can be re-written as:
\begin{equation}
	F_{\bar{\mathcal{V}}} = \dfrac{1}{2} - \dfrac{1}{\pi}\int_{0}^{\infty} \sin\left(\dfrac{\sqrt{P_o}}{2}\mathcal{H}\omega\right)\Phi_{\bar{\mathcal{V}}}\dfrac{d\omega}{\omega},
\end{equation}
where $\Phi_{\bar{\mathcal{V}}}$ is the CF of $\bar{\mathcal{V}}$, and $\mathcal{H} = \Vert \mathbf{\Delta}_o\mathbf{H}_o \Vert^2$. Therefore, to calculate the PEP of GF-SCMA, the CF of $\mathcal{\bar{V}}$ should be calculated. According to the previous definition of $\mathcal{Z}$ and $\mathcal{I}$, $\Phi_{\bar{\mathcal{V}}}$ can be decomposed into $\Phi_{\bar{\mathcal{V}}} = \varphi_{\mathcal{I}}\varphi_{\mathcal{Z}}$, where $\varphi_{\mathcal{Z}}$ and $\varphi_{\mathcal{I}}$ are the CFs of RV $\mathcal{Z}$ and $\mathcal{I}$, respectively. It should be noted that $\mathcal{Z}$ is also a complex Gaussian distributed RV, and thus $\varphi_{\mathcal{Z}} = \exp \left\{ -\frac{1}{4}\omega^2\sigma^2\mathcal{H}\right\} \label{phi_z}$. The following Lemma 2 will calculate $\varphi_\mathcal{I}$. 

\noindent \textbf{Lemma 2.} \emph{It is assumed that the interferes in a GF-SCMA network are of the same pilot, and at most one UE is transmitting data in the GF-SCMA manner with the same codebook. Moreover, it is assumed that the power of each codeword has been normalized to unit, then the CF of $\varphi_\mathcal{I}$ is given by }
\begin{equation}
\varphi_\mathcal{I} = \exp\left\{-C\left[{}_{1}F_{1}\left(-\dfrac{1}{b};1-\dfrac{1}{b};-\dfrac{v}{4}
\right)-1\right]\right\}
\end{equation}
\emph{where $C$ is a constant that can be expressed as}
\begin{equation}
C = \dfrac{\gamma\left(2,\pi\lambda_b\left(\dfrac{\rho_m}{\rho}\right)^{\frac{1}{b}}\right)}{1-\exp\left(-\pi\lambda_b\left(\dfrac{\rho_m}{\rho}\right)^{\frac{1}{b}}\right)},
\end{equation}
\emph{and $\gamma\left(\cdot,\cdot\right)$ is the lower incomplete gamma function.}

\emph{Proof}. See Appendix. \ref{Lemma 2}.

Note that when $\rho_m \to \infty$, we have $\gamma(2,\pi\lambda_b(\frac{\rho_m}{\rho})^{1/b}) \to 1$, such that $C \to 1$. According to \eqref{PEP}, it can be concluded that PEP of GF-SCMA is equivalent to the CDF of $\bar{\mathcal{V}}$; hence, based on Lemma 2, the APEP is given in the following theorem.

\noindent \textbf{Theorem 2.} \emph{With an optimal MLD, the APEP of UEs that use the same codebook in a uplink scenario with truncated full channel inversion power control scheme is given by \eqref{APEP}, where the signal-to-noise ratio (SNR) is $\text{SNR} = \frac{\rho}{\sigma^2}$, and}
\begin{figure*}[t!]
\begin{equation}
P(\tilde{\mathbf{c}_o} \to \mathbf{c}_o) = \dfrac{1}{2}-\dfrac{1}{2\pi}\int_{0}^{\infty}\dfrac{1}{v}\exp\left\{-\frac{1}{4}\dfrac{1}{\text{SNR}}v - C\left[{}_{1}F_{1}\left(-\dfrac{1}{b};1-\dfrac{1}{b};-\dfrac{v}{4}
\right)-1\right]\right\}\mathbb{E}\left[\sin\left(\dfrac{1}{2}v^{\frac{1}{2}}\sqrt{\mathcal{H}}\right)\right]  \text{d}v \label{APEP}
\end{equation} 
\hrulefill
\end{figure*}


\begin{equation}
\begin{aligned}
&\mathbb{E}\left[\sin\left(\dfrac{1}{2}v^{\frac{1}{2}}\sqrt{\mathcal{H}}\right)\right]=   \\
&\quad\quad\dfrac{\Vert \mathbf{\Delta}_o \Vert\Gamma\left(d_s + \frac{1}{2}\right)}{2\Gamma\left(d_s\right)} v^{\frac{1}{2}}
{}_{1}F_{1}\left(d_s + \dfrac{1}{2};  \frac{3}{2} ; \frac{-\Vert \mathbf{\Delta}_o \Vert^2 v }{16}\right)\label{expectation_H_R_o}.
\end{aligned}
\end{equation}
\emph{where $\Gamma\left(\cdot\right)$ denotes the gamma funtion.}

\emph{Proof}. See Appendix. \ref{Theorem 2}.

According to the APEP result in Theorem 2, we can have the following remark:

\emph{Remark 1.} The APEP decreases by increasing $\lambda_b$. This property can be simply explained by \eqref{APEP}. As $C$ is determined by $\lambda_b$; hence, APEP decreases exponentially with $\lambda_b$ according to \eqref{APEP}. Similarly, it can be also inferred that the APEP decreases by increasing $\rho_m$.

So far, the APEP of GF-SCMA system under codebook collision is obtained, in the sequel, the error rate performance will be presented based on APEP analysis.

\subsection{Approximation of the Average Symbol Error Probability} \label{Section. IV-C}
To compute ASEP from APEP, a union bound on the ASEP is given by \cite{2018Performance}
\begin{equation}
	P_{\text{ASEP}} \leq \dfrac{1}{M}\sum_{\mathbf{c}_o}\sum_{\mathbf{c}_o \neq \tilde{\mathbf{c}_o}} 	P(\tilde{\mathbf{c}_o} \to \mathbf{c}_o),  \label{union bound}
\end{equation}
where $M$ is the cardinality of $\mathcal{C}$. Note that $M^2$ PEPs should be calculated according to \eqref{union bound}, and thus the computational complexity will be high especially when $M$ is large. In this paper, to provide a simpler to compute formulation with high accuracy, we utilize an approximation under the equal-probable transmission codeword assumption \cite{2018Performance}. The approximation follows from the fact that the nearest neighbors among the codeword pair $(\tilde{\mathbf{c}_o} ,\mathbf{c}_o)$ dominate the error rate in high SNR region. So, to reduce the number of PEPs that should be calculated, we can only consider the nearest neighbors of each codeword with minimum Euclidean distance, which has main contribution to the ASEP performance. Accordingly, the ASEP of GF-SCMA can be formulated as
\begin{equation}
	P_{\text{ASEP}} \approx \dfrac{1}{M}\sum_{\mathbf{c}_o} \mathcal{N}_{\mathbf{c}_o} P(\tilde{\mathbf{c}_o} \to \mathbf{c}_o)\mid_{\mathbf{\Delta}_{\min}}, 
\end{equation}
where $\Vert\mathbf{\Delta}_{\min} \Vert^2= \min_{\mathbf{c}_o,\tilde{\mathbf{c}_o} \in \mathcal{C}}\{\Vert\mathbf{c}_o - \tilde{\mathbf{c}_o}\Vert^2, \mathbf{c}_o\neq \tilde{\mathbf{c}_o}\}$ represents the minimum Euclidean distance among all codeword pairs $(\tilde{\mathbf{c}_o} ,\mathbf{c}_o)$, and $\mathcal{N}_{\mathbf{c}_o}$ is the number of the nearest neighbors of $\mathbf{c}_o$. $P(\tilde{\mathbf{c}_o} \to \mathbf{c}_o)\mid_{\mathbf{\Delta}_{\min}}$ denotes the PEP by replacing $\mathbf{\Delta}_{\min}$ with $\mathbf{\Delta}_o$. The accuracy of such an approximation will be verified in Section. \ref{Section_V}. 

To elaborate further, some ASEP trends can be obtained based on the analysis, and thus we have the following two remarks. We have to point out that these two remarks are based on the assumption that $\text{SNR} \to \infty$ and $\rho_m \to \infty$. This indicates that an interference-limited and full channel inversion power control is considered in the following discussions.

 \emph{Remark 2.} When $\text{SNR} \to \infty$, the ASEP is independent of $\rho$. It can be observed from \eqref{APEP} that if $\text{SNR} \to \infty$, then $\frac{v}{4\text{SNR} }\to 0$, and $C \to 1$. Thereby, the ASEP is only determined by \eqref{expectation_H_R_o} and path-loss exponent $\eta$. This property implies that an error-floor emerges at very high SNR region when full channel inversion power control without truncation is utilized.

\emph{Remark 3.} Higher $b$ (or equivalently $\eta$) leads to a worse error rate performance. To explain this property, the monotonicity of ${}_{1}F_{1}(-1/b;1-1/b;-v/4)$ should be investigated. As can be seen in \eqref{phi_i_deduce}, $(-1/b)_q/(1-1/b)_q$ reflects the impact of $b$ on ${}_{1}F_{1}(-1/b;1-1/b;-v/4)$, in which $(-1/b)_q/(1-1/b)_q$ can be further simplified as $1/(1-qb)$, which follows from $\Gamma(z+1) = z\Gamma(z)$. As $q$ is a non-negative integer, and thus $(-1/b)_q/(1-1/b)_q$ is proportional to $-1/b$.

\section{Numerical Results and Discussion} \label{Simulation}
In this section, we verify our analytical results of upink GF-SCMA transmission via Monte Carlo simulation. A simulation area of 100$\text{km}^2$ is considered for the deployment of BSs and UEs. Each UE associated with its nearest BS, and employs the truncated full channel inversion power control scheme with a maximum transmit power $\rho_m = 1$ W, which is the same as \cite{2014ElSawy}. Moreover, we assume that the density of BS is $\lambda_b = 1\times 10^{-5}$, and the activation probability of UE is set as $p_a = 0.1$. Furthermore, the dense and sparse codebooks used in this paper follows the design rule in and \cite{liu2020sparse} \cite{Taherzadeh2014SCMA}, respectively. For brevity, ``Ana.'' and ``Sim'' are utilized as the abbreviation of ``Analytical'' and ``Simulation'' to represent the results derived from the theoretical analysis and numerical simulation, respectively. Without specific clarification, $\rho = -100$ dBm, $\sigma^2 = -90$ dBm, $\eta$ = 4, and $L = 6$, $K= 4$, $M = 4$, $T = 4$ for both SCMA and its variant, i.e., DCMA\footnote{As DCMA is a variant of SCMA, in which a dense codebook is utilized; hence, to avoid ambiguity, DCMA with GF transmission is also termed as GF-SCMA in this paper. Note that these two schemes are mainly distinguished by sparse degree $d_s$ of codebook.}. Remind that $\lambda_u$ in this paper denotes the intensity of UE using the same pilots. 

\begin{figure}[t!]
	\centering	
	\renewcommand{\captionfont}{\small } \renewcommand{\captionlabelfont}{\small} \centering \includegraphics[height=2.8in, width=3.2in]{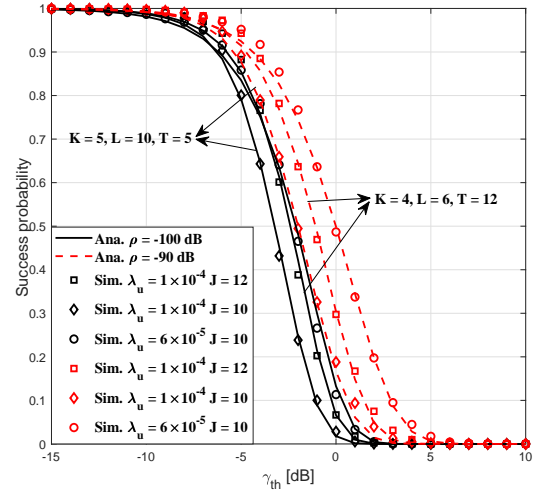}\caption{Comparing the success probability of 	GF-SCMA under different $\lambda_u$, overloading factors, and transmit power. }
	\label{fig_success_diff_dense_codebook_P}	
\end{figure}
Fig. \ref{fig_success_diff_dense_codebook_P} verifies the feasibility of the theoretical success probability in Eq. \eqref{Theo_1}. We present two scenarios for $J = 10$ and $J = 12$ with $d_s = 2$ in the figure. As shown in the figure, the simulation and analytical results derived from our proposed model match closely under different parameter setups. From Fig. \ref{fig_success_diff_dense_codebook_P}, the following observations can be obtained: 1) GF-SCMA system with less codebooks results in a lower success probability under the same $\lambda_u$ and $\gamma_{\text{th}}$. This follows from the fact that more UEs can be allocated with a codeword when $J$ is high, and thus the intra-cell interferences can be reduced; 2) The improvement of success probability by increasing the number of available codewords shrinks as the intensity of UE reduces. This is because each UE has a higher probability to possess a specific codebook in the typical cell when $\lambda_u$ is relatively low, and thus the intra-cell can be reduced; 3) The pace of deterioration of success probability when increasing the number of UE slows down as the transmit power reduces. This implies that low $\lambda_u$ benefits more from transmitting with higher power. This can be simply explained by the fact that when $\lambda_u$ decreases, the intra-cell interference is reduced. Thus, a higher $\rho$ can improve the success probability significantly for low $\lambda_u$; 4) The performance gain obtained by increasing the number of codebooks under different transmit powers are approximately the same. This indicates that the overloading factor does not have remarkable effect on success probability.

\begin{figure}[t!]
	\centering	
	\renewcommand{\captionfont}{\small } \renewcommand{\captionlabelfont}{\small} \centering \includegraphics[height=2.8in, width=3.2in]{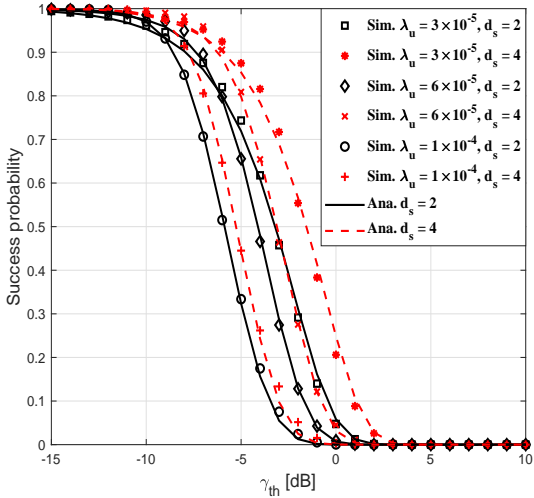}\caption{Success probability of GF-SCMA under different $\lambda_u$.}
	\label{fig_success_diff_dense}	
\end{figure}
Fig. \ref{fig_success_diff_dense} plots the success probability comparison between GF-SCMA with sparse and dense codebook. This enables us to study the influence of sparsity degree $d_s$ on the performance of GF transmission.
It is presented in Fig. \ref{fig_success_diff_dense} that dense codebook is able to achieve higher success probability. Furthermore, less performance gain can be attained when the intensity of UE goes up. It should be noted that these observations have been well predicted in Corollary 1. The other important observation is that using dense codebook can mitigate the degradation of $P_{\text{suc}}$ brings from the increase of $\lambda_u$. Thus , for GF-SCMA system, instead of assigning more distinguished codebooks to UEs, making the codeword denser is also a feasible method to cope with massive connections in practical scenarios.  However, it is worth noting that the utilization of dense codebook significantly increase the computational complexity at the receiver side as $K$ and $L$ increase \cite{liu2020sparse}. 
We also notice that for $d_s = 2$, when $\gamma_{\text{th}} < -9.5$ dB and $\gamma_{\text{th}} < -6.5$ dB, the success probability of GF-SCMA with $\lambda_u = 3\times10^{-5}$ is significantly worse than that with $\lambda_u = 1\times10^{-4}$ and $\lambda_u = 6\times10^{-5}$, respectively. A similar  phenomenon can be observed for $d_s = 4$ in the figure. For this observation, we can see the well match between the simulation results and the conclusion in Corollary 3, and the reason of such phenomenon has been explained. 
The results in Figs. \ref{fig_success_diff_dense_codebook_P} and \ref{fig_success_diff_dense} also reveal that the quadratic approximation of ${}_{2}F_{1}$ is useful in some semi-quantitative analysis of the proposed GF-SCMA analytical model\footnote{The quadratic approximation of ${}_{2}F_{1}$ is presented in Appendix C to prove Corollary 1.}.

\begin{figure}[t!]
	\centering	
	\renewcommand{\captionfont}{\small } \renewcommand{\captionlabelfont}{\small} \centering \includegraphics[height=2.8in, width=3.2in]{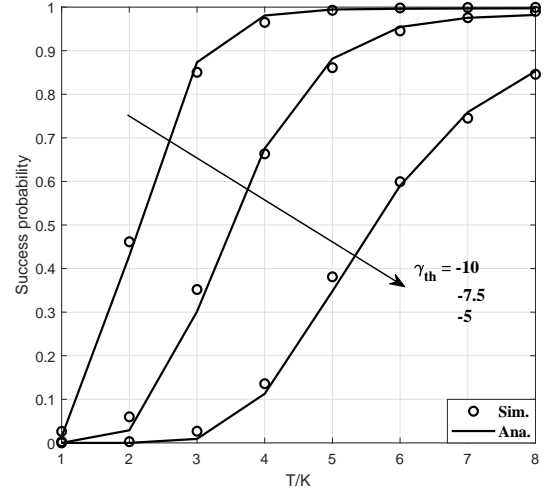}\caption{Success probability of SCMA against $T/K$. The number of codebook $J$ is determined by $T$ according to $J = TL/K$, and the results are obtained under the condition that $\lambda_u/\lambda_b = 60.$}
	\label{fig_success_tones}	
\end{figure}
Fig. \ref{fig_success_tones} plots the success probability versus the $T/K$. As such, the impact of the number of codebook, i.e., $J$, on success probability can be investigated. As can be observed from Fig. \ref{fig_success_tones}, in the relatively low \text{SINR} threshold region ($\gamma_{\text{th}}$ = -10 dB), $P_{\text{suc}}$ increases rapidly as $T/K$ increases. However, as $\gamma_{\text{th}}$ increaes, the increase of success probability by adding more available codebooks becomes relatively slow, which indicates that the slope of the curve becomes small. Therefore, it can be inferred that lower $\gamma_{\text{th}}$ is more sensitive to the growth of $J$. In this regard, the success probability performance can be significantly improved by using more OFDMA tones if $\gamma_{\text{th}}$ is relatively low. Furthermore, when $T/K$ becomes larger, i.e., more codebook can be used in the network, $P_{\text{suc}}$ gradually converges to a fixed value, and thus the increase of $J$ imposes a marginal impact on the success probability. This property is well predicted in Corollary 2 and can be explained by the fact that if $J$ is large enough under the condition that $\lambda_u$ is fixed, then $J \gg \vert 
\mathcal{U}\vert$ holds. In this regard, the number of codebook is sufficient to support all UEs in the typical cell, and thus
the aggregate intra-cell interference diminishes, leaving the GF-SCMA UEs in a typical cell approach to a fixed success probability.

\begin{figure}[t!]
	\centering	
	\renewcommand{\captionfont}{\small } \renewcommand{\captionlabelfont}{\small} \centering \includegraphics[height=2.8in, width=3.2in]{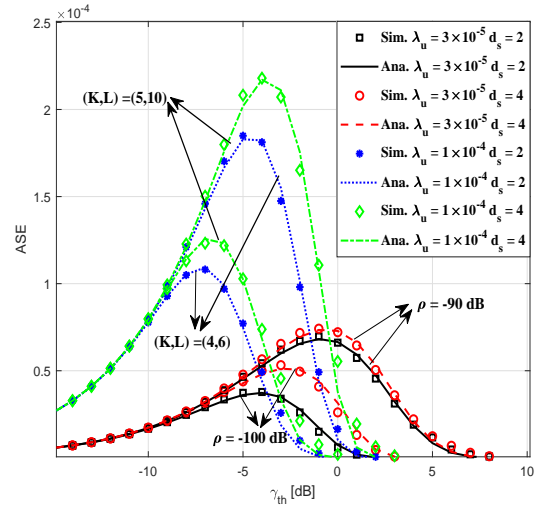}\caption{Comparison of the area spectral efficiency results under different parameter setups.}
	\label{fig_ASE}	
\end{figure}
Fig. \ref{fig_ASE} plots the ASE versus the $\text{SINR}$ threshold for different $\lambda_u$, transmit power, and the number of codebook. It can be observed from the figure that: 1) For $\lambda_u = 3\times 10^{-5}$, better ASE performance can be achieved by GF-SCMA with higher transmit power when $\gamma_{\text{th}} > -9$ dB, which indicates that increasing $\rho$ is beneficial to support more UEs in the network. Nevertheless, the advantage of ASE performance, which is achieved by making codebook denser, becomes insignificant as the increase of $\rho$. To conclude, the sparse codebook is more suitable for the transmission when $\rho$ is relatively high since it is able to strike a good balance between the ASE and detection complexity; 2) The ASE performance benefits from utilizing more available codebooks, and the $\gamma_{\text{th}}$ for GF-SCMA system with $K = 5, L = 10$ that can achieve the maximum ASE is higher than that of GF-SCMA with $K = 4, L = 6$, where the target $\text{SINR}$ threshold is approximate $\gamma_{\text{th}} = 7.5$ dB and $\gamma_{\text{th}} = 5$ dB, respectively; 3) As shown in the figure, GF-SCMA system with higher $\lambda_u$ and $J$ is able to support more UEs. However, the ASE of such GF-SCMA network suffers from a rapid decline as the increase of $\gamma_{\text{th}}$. This means that the advantage on the ASE of GF-SCMA with higher $\lambda_u$ and $J$ vanishes in the case when $\gamma_{\text{th}}$ is relatively high. Therefore, in a IoT network with dense UE deployment, GF-SCMA with less available codebooks performs relatively more robust when $\gamma_{\text{th}}$ varies. 

\begin{figure}[t!]
	\centering	
	\renewcommand{\captionfont}{\small } \renewcommand{\captionlabelfont}{\small} \centering \includegraphics[height=2.8in, width=3.6in]{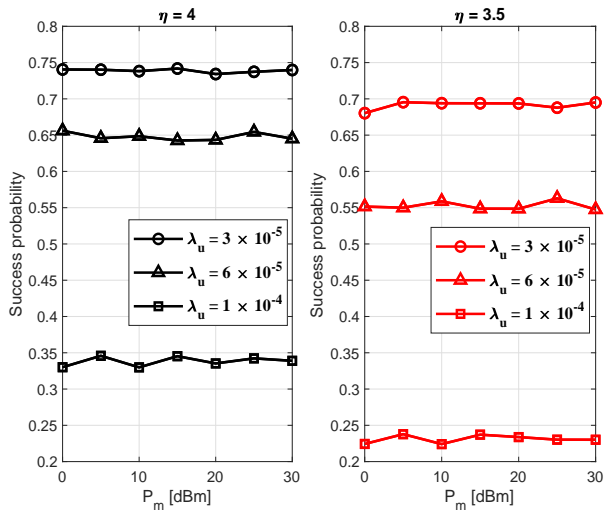}\caption{Success probability of GF-SCMA system versus maximum transmit power at $\gamma_{\text{th}} = 5$ dB.}
	\label{fig_Success_rate_Pmax}	
\end{figure}
Fig. \ref{fig_Success_rate_Pmax} plots the variation of $P_{\text{suc}}$ with $\rho_m$ for different densities and path-loss exponents $\eta$. Since the truncated channel inversion power control scheme is considered in this work, it is necessary to explore the impact of $\rho_m$ on success probability. 
As can be seen from the figure, the success probability does not vary with the maximum transmission power in a wide range, i.e., 0 to 30 dBm. This can be justified by noting the fact that decreasing  $\rho_m$ is equivalent to the increase of $\lambda_u$. As $\lambda_u$ increases, the distance between UE and its serving BS shrinks, leading to the transmission power required to satisfy the power control truncation threshold becomes smaller. In this regard, the variation of $\rho_m$ imposes a marginal effect on the success probability. Therefore, the application of such power control scheme can save cost with negligible performance loss. It is noteworthy that this property is meaningful since the practical massive IoT networks require low cost and long battery life. Furthermore, we can also observe that a lower path-loss exponent leads to the deterioration of success probability.

\begin{figure}[t!]
	\centering 
	\renewcommand{\captionfont}{\small } \renewcommand{\captionlabelfont}{\small} \centering 	
	\subcaptionbox{\label{fig_error_4point}}	
	{\includegraphics[height=2.8in, width=3.2in]{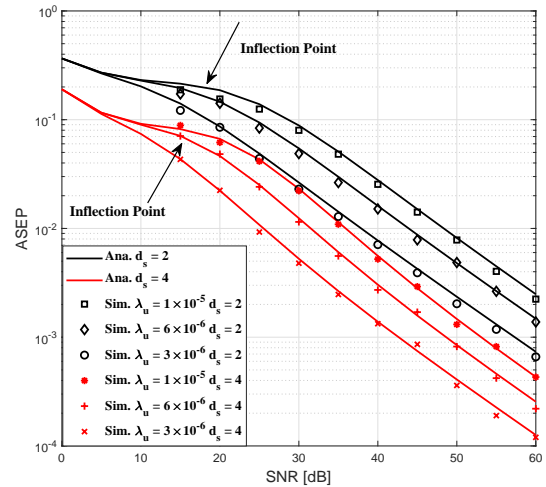}} 	
	\subcaptionbox{\label{fig_error_8point}}	
	{\includegraphics[height=2.8in, width=3.2in]{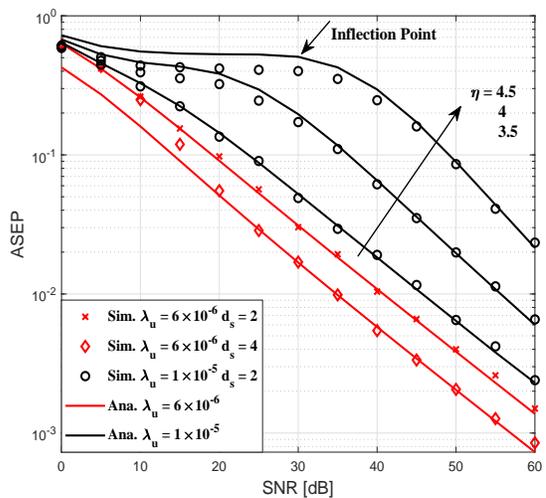}}	
	\caption{Comparisons of ASEP performance.\label{fig_error} (a) ASEP of 4-point GF-SCMA with different $\lambda_u$ and $d_s$ against SNR. (b) ASEP of 8-point GF-SCMA against SNR for different path-loss exponents.}		
\end{figure}
Figs. \ref{fig_error_4point} and \ref{fig_error_8point} plot the ASEP of 4-point and 8-point GF-SCMA system with different parameter setups, including different $\lambda_u$, $\eta$, $d_s$ and $M$. It can be firstly seen from the figures that the simulated ASEP curves can be well-approximated by the analytical results derived from Theorem 2, especially in the region of high SNR. This follows from the fact that the union bound given in \eqref{union bound} becomes more accurate at high SNR. In addition, compared with the conventional SCMA, the utilization of dense codebook shows advantage on ASEP under any $\lambda_u$, which is consistent with the results presented in \cite{liu2020sparse}. 
It should be noted that the ASEP gain that 8-point GF-SCMA with $d_s = 4$ can achieve is less than that of 4-point system. Specifically, as for GF-SCMA with $d_s = 4$ and $\lambda_u = 6 \times 10^{-6}$, it is shown in Fig. \ref{fig_error} that the performance gain of 8-point and 4-point GF-SCMA with dense codebook is 5 dB and 12 dB at ASEP=$10^{-2}$, respectively. Figs. \ref{fig_error_8point} also shows the impact that $\eta$ has on the ASEP, and the performance trends are in agreement with the result in Remark 3, i.e., the ASEP become worse as the decrease of $\eta$.

Moreover, from both Fig. \ref{fig_error_4point} and Fig. \ref{fig_error_8point}, we can observe that there is an inflection point at the low SNR region for both 4- and 8-point GF-SCMA systems. The appearance of such an inflection point can be explained by the utilization of truncated full channel inversion power control. Taking the 4-point SCMA with $d_s = 2$ and $\lambda_u = 1 \times 10^{-5}$ as an example. According to  Remark 2, if full channel inversion power control without truncation is utilized, an error floor should appear. That is to say, the ASEP does not decline as SNR increases, but keeps stable at approximate ASEP = 0.4, which can be directly spotted from Fig. \ref{fig_error_4point}. The reason for such a trend is mainly because the aggregate power of interference is comparable with the power of typical UE. However, due to the truncated power control scheme is used; hence, $\rho$ increases as the growth of SNR. This results in the power outage of more UEs in the cell, such that the error floor vanishes. 

\begin{figure}[t!]
	\centering 
	\renewcommand{\captionfont}{\small } \renewcommand{\captionlabelfont}{\small} \centering 	
	\subcaptionbox{$M=4, \text{SNR} = 30$ dB.\label{fig_error_Pmax_4point}}	
	{\includegraphics[height=2.8in, width=3.2in]{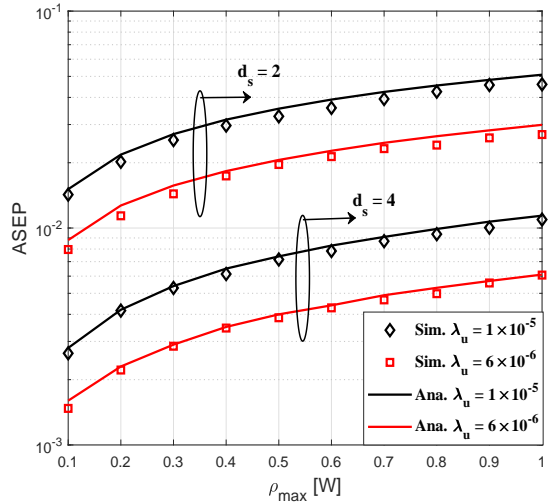}} 	
	\subcaptionbox{$M=8, \text{SNR} = 35$ dB.\label{fig_error_Pmax_8point}}	
	{\includegraphics[height=2.8in, width=3.2in]{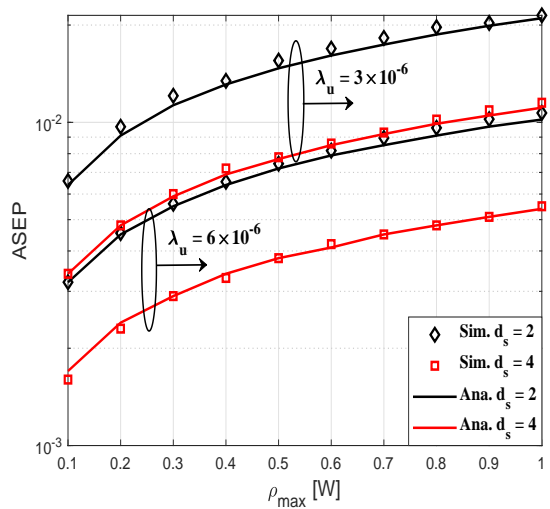}}	
	\caption{ASEP comparison of SCMA and DCMA with $M=8$ under different $\rho_m$.\label{fig_error_Pmax}}		
\end{figure}
In Figs. \ref{fig_error_Pmax_4point} and \ref{fig_error_Pmax_8point}, the impact of $\rho_m$ on ASEP is examined. In these two figures, it is shown that lower $\rho_m$ leads to better ASEP performance since more UEs with the same CTU experience power outage when $\rho_m$ is low, and thus fewer UEs undergo codebook collision. This leads to the improvement of ASEP performance when $\rho_m$ is relatively low. As $\rho_m$ becomes larger, the ASEP suffers from performance degradation, which can be attributed to the reason that more active UEs use the same CTU transmit data simultaneously because less power outage occurs. Furthermore, it can be witnessed in Fig. \ref{fig_error_Pmax} that GF-SCMA with dense codebook exhibits a similar trend as that with sparse codebook. In addition, as shown in Fig. \ref{fig_error_Pmax_8point}, for $M = 8$ or higher, dense codebook can sinificantly mitigate the deterioration of error rate performance caused by deploying more UEs in the GF-SCMA network. It is shown in the figure that GF-SCMA with $d_s = 4, \lambda_u = 6 \times 10^{-6}$ has comparable performance as GF-SCMA with $d_s = 2, \lambda_u = 3 \times 10^{-6}$ whatever $\rho_{m}$ varies.

\label{Section_V}
\section{Conclusion}
In this paper, we analyzed the success probability and error rate performance of GF-SCMA system in massive IoT networks. Furthermore, the success probability of GF-SCMA was derived by using the Gil-Paleaz inversion theorem, in which the property of MPA and SCMA codeword structure are taking into account. After that, to investigate the impact of codebook collision on reliability, the ASEP performance of GF-SCMA was studied.
Numerical results verified the accuracy of our analytical results, and showed that both success probability and ASE of GF-SCMA benefit from utilizing more available codebooks in a massive connected scenario. Moreover, it was also shown that GF-SCMA with dense codebook has superiority over GF-SCMA with sparse codebook in terms of success probability, ASE, and ASEP performance. However, the advantage on success probability by utilizing denser codebook gradually diminishes as further increasing the UE intensity and sparse degree. In a word, GF-SCMA with more candidate codebooks and higher sparse degree showed advantage in massive IoT networks at the cost of higher computational complexity overhead. Last but not least, the analytical model developed in this paper can also be applied to the analysis of other GF-CD-NOMA schemes, especially for the short spreading sequence based CD-NOMA. 

\appendices
\section{Proof of Lemma 1} \label{Lemma 1}
According to the definition of CF and the depicted system model in this paper, $\Phi_{\mathcal{I}_{inter}}$ can be formulated as
\begin{equation}
\begin{aligned}
&\Phi_{\mathcal{I}_{inter}} = \mathbb{E}\left[\exp\left\{j\omega\mathcal{I}_{inter}\right\}\right]  \\
&\stackrel{(a)}{=}\exp\left(-2\pi\dfrac{\mathcal{O}_p\lambda_u}{\lambda_b}\mathbb{E}\left[\int_{R_i}^{\infty}\left(1-\exp\left(j\omega P_i\mathcal{G}_ix^{-\eta}\right)\right)x\text{d}x\right]\right)\\
&\stackrel{(b)}{=} \exp\left(\pi\dfrac{\mathcal{O}_p\lambda_u}{\lambda_b}\left(1-{}_{2}F_{1}\left(\dfrac{-1}{b},d_s;1-\dfrac{1}{b};j\omega\rho \right)\right)\mathbb{E}\left[R_i^2\right]\right),
\end{aligned}
\end{equation}
where ($\mathnormal{a}$) holds because of the probability generating functional (PGFL) of HPPP, and $\mathcal{O}_p = \exp(-\pi\lambda_b(\rho_m/\rho)^{(1/b)})$ is the power outage probability, which has given in \cite{2014ElSawy}. ($\mathnormal{b}$) can be obtained by utilizing the results in \cite[Eq. (47)]{2019Non} and Eq. (7.621.4) in \cite{2007Gradshteyn}. After calculating the second moment of $R_i$, where $R_i \leq (\rho_m/\rho)^{(1/\eta)}$, and defining $\beta = \mathcal{O}_p\lambda_u/c\lambda_b$, \eqref{CF_inter} can be attained.

As for $\Phi_{\mathcal{I}_{intra}}$, we have \eqref{phi_intra_deduce}, where $\mathbb{P}\left\{\mathcal{U}_{in}\right\}$ is given by \cite[Eq. (10)]{Nan2018Random}. For \eqref{phi_intra_deduce},  ($\mathnormal{a}$) follows under the constraint that MPA can detect the signals that are superimposed by $J$ UEs at most, and the assumption that $\vert\mathcal{U}\vert$ UEs are of different CTUs. Thus, the intra-cell interference comes from the remaining UEs in the typical cell;  ($\mathnormal{b}$) holds because
\begin{equation}
\begin{aligned}
\sum_{u=0}^{\infty}\mathbb{P}\left\{\mathcal{U} = u\right\}&\dfrac{1}{\left(1 - j\omega\rho\right)^{d_su}} \\
&= \left[1+\beta \left(1-\left(1 - j\omega\rho\right)^{-d_s}\right)\right]^{-c-1},
\end{aligned}
\end{equation} 
 which can be proved by using the recurrence formula of gamma function and the power series of $(1-z)^{-\xi}$. After some manipulations, we have \eqref{CF_intra}. This completes the proof.
\begin{figure*}
\begin{equation}
\begin{aligned}
\Phi_{\mathcal{I}_{intra}}& = \sum_{u=0}^{\infty}\mathbb{P}\left\{\left\vert\mathcal{U}_{in}\right\vert = u\right\}\left(\mathbb{E}\left[\exp\left(j\omega\mathcal{I}_{intra}\right)\right] | \left\vert\mathcal{U}_{in}\right\vert = u\right)\\
&\stackrel{(a)}{=}\sum_{u=0}^{J-1}\mathbb{P}\left\{\left\vert\mathcal{U}_{in}\right\vert = u\right\} + \left(1 - j\omega\rho\right)^{d_s(J-1)}\sum_{u=J}^{\infty}\mathbb{P}\left\{\left\vert\mathcal{U}_{in}\right\vert = u\right\} \left(1 - j\omega\rho\right)^{-d_su}\\
&\stackrel{(b)}{=} \sum_{u=0}^{J-1}\mathbb{P}\left\{\left\vert\mathcal{U}_{in}\right\vert = u\right\} +  \left(1 - j\omega\rho\right)^{d_s(J-1)} \left[\left(1+\beta\left(1-\left(1 - j\omega\rho\right)^{-d_s}\right)\right)^{(-c-1)} - \sum_{u=0}^{J-1}\mathbb{P}\left\{\left\vert\mathcal{U}_{in}\right\vert = u\right\}\left(1 - j\omega\rho\right)^{-d_su}\right], \label{phi_intra_deduce}
\end{aligned}
\end{equation}
\hrulefill
\end{figure*}
 
\section{Proof of Theorem 1} \label{Theorem 1}
According to the definition of  success probability and Gil-Paleaz inversion theorem, $P_{\text{suc}}$ can be derived as
\begin{equation}
\begin{aligned}
&P_{\text{suc}} = \text{Pr}\left(\text{SINR}\geq \gamma_{\text{th}}\right) \\
 &=\mathbb{E}\left[\text{Pr}\left(\mathcal{I}_{inter}+\mathcal{I}_{intra}\leq\dfrac{\sum_{j\in\mathcal{U}}\rho\mathcal{G}_j}{\gamma_{\text{th}}}-\sigma^2\right)\right]\\
& = \dfrac{1}{2} - \dfrac{1}{\pi}\int_{0}^{\infty} \Im\Bigg\{ \Phi_{\mathcal{I}}\mathbb{E}\Bigg[\exp\Bigg(-j\omega\\
 &\quad\quad\quad\quad\quad\quad\quad\quad \times \left(\dfrac{1}{\gamma_{\text{th}}}\sum\limits_{j\in\mathcal{U}}\rho\mathcal{G}_j-\sigma^2\right)\Bigg)\Bigg]\Bigg\}\dfrac{\text{d}\omega}{\omega}, \label{P_success}
\end{aligned}
\end{equation}
where $\mathcal{I} = \mathcal{I}_{intra} + \mathcal{I}_{inter}$, such that $\Phi_{\mathcal{I}} = \Phi_{\mathcal{I}_{intra}}\Phi_{\mathcal{I}_{inter}}$. It should be noted that
	$\mathbb{E}[\exp(-j\omega(\sum_{j\in\mathcal{U}}\rho\mathcal{G}_j/\gamma_{\text{th}}))] =  \mathbb{E}\left[\prod_{\mathcal{U}} (1+j\omega\rho/\gamma_{\text{th}})^{-d_s}\right]$ holds because $\mathcal{G}_j \sim \text{Gamma}(d_s,1)$. Similar to the proof in Appendix. \ref{Lemma 1}, according to (4), when $\left\vert\mathcal{U}\right\vert>J-1$, 
\begin{equation}
\mathbb{E}\left[\prod_{\mathcal{U}} (1+\dfrac{j\omega\rho}{\gamma_{\text{th}}})^{-d_s}\right] = \mathbb{P}\left\{\mathcal{U}\geq J\right\} \left(1+\dfrac{j\omega\rho}{\gamma_{\text{th}}}\right)^{-Jd_s} \label{s_omega_2}
\end{equation}	
while $\left\vert\mathcal{U}\right\vert\leq J-1$, we have
\begin{equation}
\mathbb{E}\left[\prod_{\mathcal{U}} (1+\dfrac{j\omega\rho}{\gamma_{\text{th}}})^{-d_s}\right] = \sum_{u = 0}^{J-1}\mathbb{P}\left\{\mathcal{U} =  u\right\} \left(1+\dfrac{j\omega\rho}{\gamma_{\text{th}}}\right)^{-d_s(u+1)} \label{s_omega_1}
\end{equation}	
By substituting \eqref{s_omega_1} and \eqref{s_omega_2} into \eqref{P_success}, and after some algebraic manipulations, the theorem can be proved.
\section{Proof of Corollary 1} \label{Corollary 1}
Since the number of codebook is large enough to support all the intra-cell UEs, hence, $\Phi_{\mathcal{I}_{intra}} = 0$, and $\mathcal{S}_{\omega} = [1+\beta (1-(1 + j\omega\rho/\gamma_{\text{th}})^{-d_s})]^{-c-1}$. Given that $\rho$ usually takes small values, and thus $\tilde{\mathcal{S}_{\omega}}$ can be approximated as $\tilde{\mathcal{S}_{\omega}} \approx \left(1 - j\left(c+1\right)d_s\beta\rho\omega/\gamma_{\text{th}}\right)$ by using two times Laurent expansion. Then, we use
\begin{equation}
{}_{2}F_{1}\left(-\dfrac{1}{b},d_s;1-\dfrac{1}{b};j\omega\rho \right) \approx 1 - \dfrac{j\rho d_s}{b-1}\omega + \dfrac{d_s\left(d_s+1\right)\rho^2}{4b-2}\omega^2, \label{appro_2F1}
\end{equation}
which follows from the series representation of hypergeometric function, and the reason that we use quadratic approximation is to improve the accuracy since $d_s > 1$. After substituting \eqref{appro_2F1} and $\mathcal{S}_{\omega}$ into \eqref{Theo_1}, $P_{\text{suc}}$ can be formulated as:
\begin{equation}
\begin{aligned}
P_{\text{suc}} &\stackrel{(a)}{\approx} \dfrac{1}{2} - \dfrac{1}{\pi}\int_{0}^{\infty}\Bigg\{\dfrac{\sin\left(\dfrac{\beta cd_s\rho}{b-1}\omega\right)}{\omega} - \dfrac{\beta\left(c+1\right)d_s\rho}{\gamma_{\text{th}}} \\
&\quad\times \cos\left(\dfrac{\beta cd_s\rho}{b-1}\omega\right)\exp\left(-\dfrac{d_s\left(d_s+1\right)\rho^2}{4b-2}\omega^2\right)\Bigg\}\text{d}\omega \\
&\stackrel{(b)}{=} \dfrac{1}{2} - \frac{1}{\sqrt{\pi}}\Bigg\{\beta\sqrt{4b-2}\left(\dfrac{1}{1 + \frac{1}{d_s}}\right)^{\frac{1}{2}}\Bigg[\dfrac{c}{2\left(b-1\right)}\\
&\quad -\dfrac{c+1}{\gamma_{\text{th}}}\exp\left(-\dfrac{\lambda_u^2\left(2b-1\right)}{2\lambda_b^2\left(b-1\right)^2}\left(\dfrac{1}{1 + \frac{1}{d_s}}\right)\right)\Bigg]\Bigg\} \label{Success_appr}
\end{aligned}
\end{equation}
where ($\mathnormal{a}$) follows by substituting \eqref{appro_2F1} and $\tilde{\mathcal{S}_{\omega}}$ into \eqref{Theo_1}, and the result in  ($\mathnormal{b}$)  can be obtained by solving the integral with the notable approximation $\sin x \approx x$, and the integral results of \cite[Eq. (3.321.3)]{2007Gradshteyn} and \cite[Eq. (3.896.4)]{2007Gradshteyn}. After some algebraic manipulations, the conclusion can be proved.

\section{Proof of Lemma 2} \label{Lemma 2}
According to the definition of CF, and after some changes of variables, $\varphi_{\mathcal{I}}$ can be expressed as \eqref {phi_i}
\begin{equation}
\begin{aligned}
\varphi_{\mathcal{I}} &= \mathbb{E}\left[\exp\left(j\omega\mathcal{I}\right)\right] \\
		    &= \mathbb{E}\left[\prod\limits_{\text{UE}_i \in \Phi_U\backslash \text{UE}_o}\exp\left(j\omega\sqrt{\rho}R_i^{b}\hat{g_i}D_i^{-b}\right)\right]\\
		    &\approx \exp\left\{-2\pi\lambda_b \int_{t}^{\infty}\mathbb{E}\left[1-\exp\left(j\omega\sqrt{\rho}R_i^{b}\hat{g_i}x^{-b}\right)\right]x\text{d}x\right\}, 	\label{phi_i}
\end{aligned}
\end{equation}
where $t = R_i$, $\hat{g_i} = \mathbf{c}_i\mathbf{H}_i\left(\mathbf{\Delta}_o\mathbf{H}_o\right)^{\dagger}$, and $\text{UE}_o$ is a UE in the typical cell. The approximation in \eqref{phi_i} results from approximating the interfering UEs as PPP so that the PGFL of PPP can be applied. Note that the lower limits of integral follows from the fact that $D_i>R_i$.
By further defining $y = x\omega^2$, we have
	\begin{equation}
	\begin{aligned}
		&\varphi_{\mathcal{I}} \\
		&= \exp\left\{\dfrac{-2\pi\lambda_b}{b} \int_{0}^{\omega R_i^{-b}} \mathbb{E}_{\hat{g_i}}\left[1-\cos\left(\sqrt{\rho}R_i^{b}\hat{g_i}y\right)y^{-\left(1+\frac{2}{b}\right)}dy\right]\right\} \\
		& \stackrel{(a)}{=} \exp \left\{-\pi\lambda_b 	\mathbb{E}\left[R_i^2
		\left[{}_{1}F_{2}\left(\frac{-1}{b} ; \frac{1}{2}, 1-\frac{1}{b} ; \frac{-\omega^{2} \rho \hat{g_i}^{2}}{4}\right)-1\right]\right]\right\}\\
		& \stackrel{(b)}{=} \exp \left\{-\pi\lambda_b \mathbb{E}\left[R_i^2\right] \sum_{q=1}^{\infty} \dfrac{\left(-\frac{1}{b}\right)_q\left(\dfrac{-\omega^2 \rho }{4}\right)^q }{q!\left(1-\frac{1}{b}\right)_q\left(\frac{1}{2}\right)_{q}} \mathbb{E}_{\hat{g_i}}\left[\hat{g_i}^{2q}\right]\right\} \\
		& \stackrel{(c)}{=} \exp \left\{-\pi\lambda_b \mathbb{E}\left[R_i^2\right] \sum_{q=1}^{\infty}\dfrac{(2q-1)!!}{2^q\left(\frac{1}{2}\right)_{q}} \dfrac{\left(-\frac{1}{b}\right)_q\left(\dfrac{-\omega^2 \rho\mathcal{H}}{4}\right)^q}{q!\left(1-\frac{1}{b}\right)_q} 
		\right\} 
		\label{phi_i_deduce}
	\end{aligned}
\end{equation}
where ($\mathnormal{a}$) follows from Eq. (3.771.4) in \cite{2007Gradshteyn}; ($\mathnormal{b}$) results from the series representation of ${}_{1}F_{2}$, where $(\cdot)_q$ is the Pochhammer symbol; ($\mathnormal{c}$) holds because the utilization of the same codebook in SCMA and the independence between $R_i$ and $\hat{g_i}$.
The rest of the proof follows from $(2q-1)!! = 2^q(\frac{1}{2})_{q}$, and the character of unimodular codeword, i.e., $\sum_{s=1}^{d_s}\vert c_s\vert^2 = 1$, such that $\mathbb{E}_{c_s}[\sum_{s=1}^{d_s}\vert c_s\vert^2]^q = 1$. By using the series representation of ${}_{1}F_{1}$, the lemma can be proved.

\section{Proof of Theorem 2} \label{Theorem 2}
To this end, APEP can be obtained by averaging over $\mathcal{H}$. As a result, we have 
\begin{equation}
	P(\tilde{\mathbf{c}_o} \to \mathbf{c}_o) = \dfrac{1}{2} - \dfrac{1}{\pi}\mathbb{E}\left[ \int_{0}^{\infty} \sin\left(\dfrac{\sqrt{P_o}}{2}\mathcal{H}\omega\right)\varphi_{\mathcal{I}}\varphi_{\mathcal{Z}} \dfrac{\text{d}\omega}{\omega} \right]. \label{PEP_exp}
\end{equation} 
 By substituting $\varphi_{\mathcal{Z}}$ and \eqref{phi_i_deduce} into \eqref{PEP_exp}, and making a change of variable $v = \rho\mathcal{H}\omega^2$, then \eqref{APEP} follows. Thereby, the rest of the proof focus on deriving a closed-form of $\mathbb{E}[\sin\left(\frac{1}{2}v^{\frac{1}{2}}\sqrt{\mathcal{H}}\right)]$. It should be noted that $\mathcal{H} = \Vert \Delta_o\mathbf{H}_o\Vert^2$, and $\sum_{s=1}^{d_s}\Delta_o^{(s)}h_o^{(s)}\sim\mathcal{CN}(0,\Vert \mathbf{\Delta}_o\Vert^2)$, hence, $\mathcal{H}\sim\text{Gamma}(d_s,\Vert \mathbf{\Delta}_o\Vert^2)$. Following from such fact, \eqref{expectation_H} can be obtained,
 	 \begin{equation}
 	\begin{aligned}
 	&\mathbb{E}\left[\sin\left(\dfrac{1}{2}v^{\frac{1}{2}}\sqrt{\mathcal{H}}\right)\right]\\
 	&\stackrel{(a)}{=} 
 	\dfrac{\Vert \mathbf{\Delta}_o \Vert^{-2d_s}}{\Gamma\left(d_s \right)} \int_{0}^{\infty}\sin\left(\dfrac{1}{2}v^{\frac{1}{2}}\sqrt{\mathcal{H}}\right) \sqrt{\mathcal{H}}^{2d_s-2}e^{-\Vert \mathbf{\Delta}_o \Vert^2 \mathcal{H}}\text{d}\mathcal{H}\\
 	&\stackrel{(b)}{=}\dfrac{\Vert \mathbf{\Delta}_o \Vert v^{\frac{1}{2}}}{2\Gamma\left(d_s \right)}  \Gamma\left(d_s+\dfrac{1}{2}\right) e^{-\frac{\Vert \mathbf{\Delta}_o \Vert^2v}{16}}  {}_{1}F_{1}\left(1-d_s;  \frac{3}{2} ; \frac{\Vert \mathbf{\Delta}_o \Vert^2v}{16}\right)\\
 	&\stackrel{(c)}{=} \Vert \mathbf{\Delta}_o \Vert \dfrac{\Gamma\left(d_s + \dfrac{1}{2}\right)}{\Gamma\left(d_s\right)}\dfrac{v^{\frac{1}{2}}}{2}
  {}_{1}F_{1}\left(d_s + \dfrac{1}{2};  \frac{3}{2} ; \frac{-\Vert \mathbf{\Delta}_o \Vert^2v}{16}\right)
 	 \label{expectation_H}
 	\end{aligned}
 	\end{equation}
where ($\mathnormal{a}$) results from the distribution of random variable $\mathcal{H}$; ($\mathnormal{b}$) holds by using the integral result \cite[Eq. (3.952.7)]{2007Gradshteyn}; ($\mathnormal{c}$) follows from ${}_{1}F_{1}(a;b;z) = \exp(z){}_{1}F_{1}(b-a;b;-z)$.  
\ifCLASSOPTIONcaptionsoff
\newpage
\fi
\bibliographystyle{IEEEtran}
\bibliography{GF_CD_NOMA_ana}

\end{document}